\documentclass[11pt]{article}

\usepackage{graphicx}
\usepackage{dcolumn}
\usepackage{bm}

\begin{document}


\title{Supersymmetric methods in the traveling variable: inside neurons and at the brain scale}

\author{H.C. Rosu\footnote{Electronic mail: hcr@ipicyt.edu.mx}
$^{\clubsuit}$,
O. Cornejo-P\'erez\footnote{Electronic mail: ocornejo@cimat.mx}
$^{\spadesuit}$, and J.E. P\'erez-Terrazas\footnote{Electronic mail:
jenrique@ipicyt.edu.mx}
$^{\clubsuit}$\\
$^{\clubsuit}$ Division of Advanced Materials, IPICyT, \\Apartado Postal 3-74, Tangamanga, 78231 San Luis Potos\'{\i}, S.L.P., M\'exico\\
$^{\spadesuit}$ Centro de Investigaci\'on en Matem\'aticas (CIMAT),\\
Apartado Postal 402, 36000 Guanajuato, Gto., M\'exico }

\maketitle

\begin{abstract}
We apply the mathematical technique of factorization of differential
operators to two different problems. First we review our results
related to the supersymmetry of the Montroll kinks moving onto the
microtubule walls as well as mentioning the sine-Gordon model for
the microtubule nonlinear excitations. Second, we find analytic
expressions for a class of one-parameter solutions of a sort of
diffusion equation of Bessel type that is obtained by supersymmetry
from the homogeneous form of a simple damped wave equations derived
in the works of P.A. Robinson and collaborators for the
corticothalamic system. We also present a possible interpretation of
the diffusion equation in the brain context.
\end{abstract}
\vskip 0.1in

\vskip 0.1in



\section{Nonlinear biological excitations}
The possibility of soliton excitations in biological structures has
been first pointed out by Englander et al \cite{E80} in 1980 who
speculated that the so-called `open states' units made of
approximately ten adjacent open pairs  in long polynucleotide double
helices could be thermally induced solitons of the double helix due
to a coherence of the twist deformation energy. Since then a
substantial amount of literature has been accumulating on the
biological significance of DNA nonlinear excitations (for a recent
paper, see \cite{bash06}). On the other hand, the idea of nonlinear
excitations has emerged in 1993 in the context of the microtubules
(MTs) \cite{mtub1}, the dimeric tubular polymers that contribute the
main part of the eukaryotic cytoskeleton. In the case of neurons,
MTs are critical for the growth and maintenance of axons. It is
known that axonal MTs are spatially organized but are not under the
influence of a MT-organizing center as in other cells. We also
remind that in 1995 Das and Schwarz have used a two-dimensional
smectic liquid crystal model to show the possibility of electrical
solitary wave propagation in cell membranes \cite{ds95}.
Nevertheless, there is no clear experimental evidence at the moment
of any of these biological solitons and kinks.

\section{Supersymmetric MT Kinks}

Based on well-established results of Collins, Blumen, Currie and
Ross \cite{1} regarding the dynamics of domain walls in
ferrodistortive materials, Tuszy\'nski and collaborators
\cite{mtub1,mtub2} considered MTs to be ferrodistortive and studied
kinks of the Montroll type \cite{mont} as excitations responsible
for the energy transfer within this highly interesting biological
context.

The Euler-Lagrange dimensionless equation of motion of
ferrodistortive domain walls as derived in \cite{1} from a
Ginzburg-Landau free energy with driven field and dissipation
included is of the travelling reaction-diffusion type
\begin{equation} \label{1}
\psi ^{''}+\rho\psi ^{'}-\psi ^3 +\psi+\sigma=0~,
\end{equation}
where the primes are derivatives with respect to a travelling
coordinate $\xi =x-vt$, $\rho$ is a friction coefficient and
$\sigma$ is related to the driven field \cite{1}.

There may be ferrodistortive domain walls that can be identified
with the Montroll kink solution of Eq.~(\ref{1})
\begin{equation}  \label{2}
M(\xi)=\alpha _1+\frac{\sqrt{2}\beta}{1+\exp(\beta\xi)}~,
\end{equation}
where $\beta=(\alpha _2-\alpha _1)/\sqrt{2}$ and the parameters
$\alpha _1$ and $\alpha _2$ are two nonequal solutions of the cubic
equation
\begin{equation} \label{3}
(\psi -\alpha _1)(\psi -\alpha _2)(\psi -\alpha _3)=\psi ^3 -\psi
-\sigma~.
\end{equation}


Rosu has noted that Montroll's kink can be written as a typical
$\tanh$ kink \cite{rosu}
\begin{equation} \label{M}
M(\xi)= \gamma -\tanh\left(\frac{\beta \xi}{2}\right)~,
\end{equation}
where $\gamma \equiv \alpha _1 +\alpha _2=1+\frac{\alpha
_1\sqrt{2}}{\beta}$.
The latter relationship allows one to use a simple construction
method of exactly soluble
double-well potentials in the Schr\"odinger equation proposed by
Caticha \cite{cat}. The scheme is a non-standard application of
Witten's supersymmetric quantum mechanics \cite{w81} having as the
essential assumption the idea of considering the $M$ kink as the
switching function between the two lowest eigenstates of the
Schr\"odinger equation with a double-well potential. Thus
\begin{equation} \label{phiM}
\phi _1=M\phi _0~,
\end{equation}
where $\phi _{0,1}$ are solutions of $\phi ^{''}_{0,1}+[\epsilon
_{0,1}-u(\xi)] \phi _{0,1}(\xi)=0$, and $u(\xi)$ is the double-well
potential to be found.

   \begin{figure}[h!]
     \centering
     \includegraphics[width=7 cm, height=7 cm]{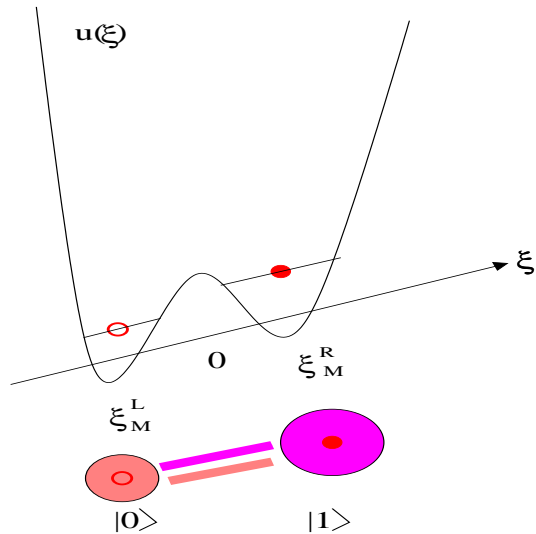}
     \caption{
     {\small Single electron within the traveling double-well potential $u(\xi)$ as a qubit. The electron can switch from one wall to
     another by tunneling and the relation
     between the wavefunctions in the two wells is given by Eq.~(\ref{phiM}).}}
     \label{dipchain}
   \end{figure}

Substituting Eq.~(\ref{phiM}) into the Schr\"odinger equation for
the subscript 1 and substracting the same equation multiplied by the
switching function for the subscript 0, one obtains
\begin{equation} \label{phiR}
\phi ^{'}_{0}+R_M\phi _0=0~,
\end{equation}
where $R_M$ is given by
\begin{equation} \label{R}
R_M=\frac{M^{''}+\epsilon M}{2M^{'}}~,
\end{equation}
and $\epsilon=\epsilon _1-\epsilon _0$ is the lowest energy
splitting in the double-well Schr\"odinger equation. In addition,
notice that Eq.~(\ref{phiR}) is the basic equation introducing the
superpotential $R$ in Witten's supersymmetric quantum mechanics,
i.e., the Riccati solution.
For Montroll's kink the corresponding Riccati solution reads
\begin{equation} \label{RM}
R_M(\xi)=-\frac{\beta}{2}{\rm
tanh}\left(\frac{\beta}{2}\xi\right)+\frac{\epsilon}{2\beta}\Bigg[\sinh(\beta\xi)+
2\gamma\cosh ^2\left(\frac{\beta}{2}\xi\right)\Bigg]
\end{equation}
and the ground-state Schr\"odinger function is found by means of
Eq.~(\ref{phiR})
\begin{eqnarray} \label{phi0}
\phi _{0,M}(\xi) &=& \phi _0(0)\cosh\left(\frac{\beta}{2}\xi\right)
\exp\left(\frac{\epsilon}{2\beta ^2}\right)
\exp\left(-\frac{\epsilon}{2\beta ^2}\Big[ \cosh (\beta \right.
\xi)\nonumber\\
&& \left. -\gamma \beta\xi-\gamma\sinh(\beta\xi)\Big]\right)~,
\end{eqnarray}
while $\phi _1$ is obtained by switching the ground-state wave
function by means of $M$. This ground-state wave function is of
supersymmetric type
\begin{equation} \label{phi}
\phi _{0,M}(\xi)=\phi _{0,M}(0)\exp\Bigg[-\int_0^{\xi}
R_M(y)dy\Bigg]~,
\end{equation}
where $\phi _{0,M}(0)$ is a normalization constant.

The Montroll double well potential is determined up to the additive
constant $\epsilon _0$ by the `bosonic' Riccati equation
\begin{eqnarray} \label{u}
u_M(\xi)&=&R_M^2-R_M^{'}+\epsilon _0 = \frac{\beta
^2}{4}+\frac{(\gamma ^2 -1)\epsilon ^2}{4\beta
^2}+\frac{\epsilon}{2}+\epsilon _0 \nonumber\\
&&+ \frac{\epsilon}{8\beta^2}\Big[\left(4\gamma ^2\epsilon
+2(\gamma^2 +1)\epsilon{\rm cosh} (\beta \xi )-8\beta^2\right){\rm
cosh} (\beta \xi ) \nonumber\\
&&-4\gamma \left(\epsilon +\epsilon {\rm cosh}(\beta \xi)-2\beta ^2
\right) {\rm sinh } (\beta \xi)\Big].
\end{eqnarray}
If, as suggested by Caticha, one chooses the ground
state energy to be
\begin{equation} \label{eps}
\epsilon _0=-\frac{\beta ^2}{4}-\frac{\epsilon}{2}+\frac{\epsilon
^2}{4\beta ^2} \left(1-\gamma ^2\right)~,
\end{equation}
then $u_M(\xi)$ turns into a travelling, asymmetric Morse
double-well potential of depths depending on the Montroll parameters
$\beta$ and $\gamma$ and the splitting $\epsilon$
\begin{equation} \label{U}
U_{0, m}^{L,R}=\beta ^2\Bigg[1\pm \frac{2\epsilon
\gamma}{(2\beta)^2}\Bigg]~,
\end{equation}
where the subscript $m$ stands for Morse and the superscripts $L$
and $R$ for left and right well, respectively. The difference in
depth, the bias, is $\Delta _m\equiv U_0^L-U_0^R=2\epsilon\gamma$,
while the location of the potential minima on the traveling axis is
at
\begin{equation}  \label{xiLR}
\xi _{m}^{L,R}=\mp\frac{1}{\beta}\ln \Bigg[\frac{(2\beta)^2\pm
2\epsilon\gamma}{\epsilon(\gamma\mp 1)}\Bigg]~,
\end{equation}
that shows that $\gamma \neq \pm 1$.
\bigskip


An extension of the previous results is possible if one notices that
$R_M$ in Eq.~(\ref{RM}) is only the particular solution of
Eq.~(\ref{u}). The general solution is a one-parameter function of
the form
\begin{equation} \label{genR}
R_M(\xi
;\lambda)=R_M(\xi)+\frac{d}{d\xi}\Big[\ln(I_M(\xi)+\lambda)\Big]
\end{equation}
and the corresponding one-parameter Montroll potential is given by
\begin{equation} \label{genu}
u_M(\xi ;\lambda)=u_M(\xi)-2\frac{d^2}{d\xi
^2}\Big[\ln(I_M(\xi)+\lambda)\Big]~.
\end{equation}
In these formulas, $I_M(\xi)=\int ^{\xi}\phi _{0,M}^2(\xi)d\xi$ and
$\lambda$ is an integration constant that is used as a deforming
parameter of the potential and is related to the irregular zero
mode. The one-parameter Darboux-deformed ground state wave function
can be shown to be
\begin{equation} \label{wfl}
\phi_{0,M}(\xi ;\lambda)=\sqrt{\lambda(\lambda+1)}\frac{\phi
_{0,M}}{I_M(\xi)+\lambda}~,
\end{equation}
where $\sqrt{\lambda(\lambda+1)}$ is the normalization factor
implying that $\lambda \notin [0,-1]$. Moreover, the 
one-parameter potentials and wave functions display singularities at
$\lambda_s=-I_M(\xi_s)$. For large values of $\pm\lambda$ the
singularity moves towards $\mp \infty$ and the potential and ground
state wave function recover the shapes of the non-parametric
potential and wave function. The one-parameter Morse case
corresponds formally to the change of subscript $M\rightarrow m$ in
Eqs.~(\ref{genR}) and (\ref{genu}).
For the single well Morse potential  the one-parameter procedure has
been studied by Filho \cite{filho} and Bentaiba et al
\cite{bentaiba}.

The one-parameter extension leads to singularities in the
double-well potential and the corresponding wave functions. If the
parameter $\lambda$ is positive the singularity is to be found on
the negative $\xi$ axis, while for negative $\lambda$ it is on the
positive side. Potentials and wave functions with singularities are
not so strange as it seems \cite{cs} and could be quite relevant
even in nanotechnology where quantum singular interactions of the
contact type are appropriate for describing nanoscale quantum
devices. We interpret the singularity as representing the effect of
an impurity moving along the MT in one direction or the other
depending on the sign of the parameter $\lambda$. The impurity may
represent a protein attached to the MT or a structural discontinuity
in the arrangement of the tubulin molecules. This interpretation of
impurities has been given by Trpi\v{s}ov\'a and Tuszy\'nski in
non-supersymmetric models of nonlinear MT excitations \cite{tt}.

\section{The sine-Gordon MT solitons}

Almost simultaneously with Sataric, Tuszynski and Zakula, there was
another group, Chou, Zhang and Maggiora \cite{czm94}, who published
a paper on the possibility of kinklike excitations of sine-Gordon
type in MTs but in a biological journal. Even more, they assumed
that the kink is excited by the energy released in the hydrolysis of
GTP $\rightarrow$ GDP in microtubular solutions. As the kink moves
forward, the individual tubulin molecules involved in the kink
undergo motion that can be likened to the dislocation of atoms
within the crystal lattice.

\medskip

They performed an energy estimation showing that a kink in the
system possesses about 0.36 - 0.44 eV, which is quite close to the
0.49 eV of energy released from the hydrolysis of GTP.

Moreover, they assumed that the interaction energy $U(r)$ between
two neighboring tubulin molecules along a protofilament is harmonic:

\begin{equation}\label{chou1}
U(r)\approx \frac{1}{2}k(r-a_0)^2~,
\end{equation}
where $k=\frac{d^2U(a_0)}{dr^2}$ and $r=x_i-x_{i-1}$. In addition to
this kind of nearest neighbor interaction, a tubulin molecule is
also subjected to interactions with the remaining tubulin molecules
of the MT, i.e., those in the same protofilament but not nearest
neighbor to it.

Chou et al cite pages 425-427 in the book of R.K. Dodd et al
(Solitons and Nonlinear Wave Equations, Academic Press 1982) for the
claiming that this interaction for the $i$th tubulin molecule of a
protofilament can be approximated by the following periodic
effective potential
\begin{equation}\label{chou2}
U_i=U_0\left(1-\cos \frac{2\pi \xi _i}{a_0}\right)~,
\end{equation}
where $U_0$ is the half-height of the potential energy barrier and
$\xi _i$ is the displacement of the $i$th tubulin molecule from the
equilibrium position within a particular protofilament.

\medskip

Introducing the new variable $\phi _i=\frac{2\pi}{a_0}\xi _i$ the
following sine-Gordon equation is obtained
\begin{equation}\label{chou3}
m\frac{\partial ^2\phi}{\partial t^2}=ka_0^2\frac{\partial
^2\phi}{\partial x^2}-\left(\frac{2\pi}{a_0}\right)^2U_0\sin \phi~
\end{equation}
that can be reduced to the standard form of the sine-Gordon equation
\begin{equation}\label{chou-sGstandard}
\frac{\partial ^2\phi}{\partial x^2}-\frac{1}{c^2}\frac{\partial
^2\phi}{\partial t^2}=\frac{1}{l^2}\sin \phi
\end{equation}
if one sets $c^2=\frac{ka_0^2}{m}$ and $l^{-2}=\frac{4\pi
^2U_0}{ka_0^4}$. Now, it is well known that the sine-Gordon equation
has the famous inverse tangent kink solution
\begin{equation}\label{chou-sGk}
\phi=\tan ^{-1}\left(\exp[\pm \frac{\gamma}{l}(x-vt)]\right)~,
\end{equation}
where $\gamma =\frac{1}{\sqrt{1-\frac{v^2}{c^2}}}$ is an acoustic
Lorentz factor and $w=\frac{\gamma}{l}$ is the kink width.

\medskip
Most interestingly, the momentum of a tubulin dimer is strongly
localized:
\begin{equation}\label{chou-mom}
p=\frac{d(m\xi)}{dt}=\frac{ma_0}{\pi}\frac{\gamma v}{l}{\rm
sech}\bigg[-\frac{\gamma}{l}(x-vt)\bigg]~.
\end{equation}
This momentum function possesses a very high and narrow peak at the
center of the kink width implying that the corresponding tubulin
molecule will have maximum momentum when it is at the top of the
periodic potential. According to Chou et al this remarkable feature
occurs only in nonlinear wave mechanics.

\bigskip


Interestingly, for purposes of illustration, these authors
have assumed the width of a kink $w\approx 3a_0$. Therefore, with
the kink moving forward, the affected region always involves three
tubulin molecules.
For a general case, however, the width $w$ of a kink can be
calculated from
\begin{equation}\label{width}
w=\frac{a_0}{2\pi}\sqrt{\frac{ka_0^2}{U_0}}~,
\end{equation}
if the force constant $k$ between two neighboring tubulin molecules
along a protofilament, the distance $a_0$ of their centers, and the
energy barrier $2U_0$ of the periodic, effective potential are
known. Then the number of tubulin molecules involved in a kink is
given by
\begin{equation}\label{no-tub}
\frac{w}{a_0}=(2\pi)^{-1}\sqrt{ka_0^2/U_0}~.
\end{equation}


It is further known that the tubulin molecules in a MT are held by
noncovalent bonds, therefore the interaction among them might
involve hydrogen bonds, van der Waals contact, salt bridges, and
hydrophobic interactions.

It was found by Israelachvili and Pashley \cite{ip82} that the
hydrophobic force law over the distance range 0-10 nm at 21$^o$C is
well described by
\begin{equation}
\frac{F_H}{R}=Ce^{-D/D_0}\, N/m~,
\end{equation}
where $D$ is the distance between tubulin molecules, $D_0$ is a
decay length, and $R=\frac{R_1R_2}{R_1+R_2}$ is a harmonic mean
radius for two hydrophobic solute molecules, all in nm. $R$ is 4 nm
in the case of tubulin.

\subsection{More on the hydrolysis and solitary waves in MTs
}

Inside the cell, the MTs exist in an unstable dynamic state
characterized by a continuous addition and dissociation of the
molecules of tubulin. The polypeptides $\alpha$ and $\beta$ tubulin
each bind one molecule of guanine nucleotide with high affinity. The
nucleotide binding site on $\alpha$ tubulin binds GTP {\em
nonexchangeably} and is referred to as the N site. The binding site
on $\beta$ tubulin {\em exchanges} rapidly with free nucleotide in
the tubulin heterodimer and is referred to as the E site.

Thus, the addition of each tubulin is accompanied by the hydrolysis
of GTP 5' bound to the $\beta$ monomer. In this reaction an amount
of energy of $6.25 \times 10^{-21}$ J is freed that can travel along
MTs  as a kinklike solitary wave.

The exchangeable GTP hydrolyses very soon after the tubulin binds to
the MT. At $pH=7$ this reaction takes place according to the
formula:
\begin{equation}\label{hydro}
GTP ^{4-}+H_2O \rightarrow GDP^{3-}+HPO_{4}^{2-}+H^{+}+\Delta _H E~.
\end{equation}

The last mathematical formulation of the manner in which the energy
$\Delta _H E$ is turned into a kink excitation claims that the
hydrolysis causes a dynamical transition in the structure of tubulin
\cite{st05}.

\section{Quantum information in the MT walls
}


Biological information processing, storage, and transduction
occurring by "computer-like" transfer and resonance among the dimer
units of MTs have been first suggested by Hamerrof and Watt
\cite{hw82} and enjoys much speculative activity \cite{faber05}.

For the case of sine-Gordon solitons, the information transport has
been investigated by Abdalla et al \cite{abdalla01}.

Recently Shi and collaborators \cite{shi06} worked out a processing
scheme of quantum information along the MT walls by using previous
hints of Lloyd for two-level pseudospin systems \cite{llo93}.  The
MT wall is treated as a chain of three types of two pseudospin-state
dimers. A set of appropriate resonant frequencies has been given.
They conclude that specific frequencies of laser pulse excitations
can be applied in order to generate quantum information processing.

Lloyd's scheme uses the driving of a quantum computer by means of a
sequence of laser pulses. He assumes a 1-dimensional arrangement of
atoms of two types (A and B) that could be each of them in one of
two states and are affected only by nearest neighbors. Then,
information processing could be performed by laser pulses of
specific frequencies $\omega _{K_{\alpha, \beta}}$,
that change the state of the atom of the K kind (A or B type) if in
a pair of atoms AB, A is in $\alpha$ state and B is in state
$\beta$.

\section{Supersymmetry at the Brain Scale}
Neuronal activity is the result of the propagation of impulses
generated at the neuron cell body and transmitted along axons to
other neurons. Recently, Robinson and collaborators \cite{rob04}
obtained simple damped wave equations for the axonal pulse fields
propagating at speed $v_a$ between two populations, $a$ and $b$, of
neurons in the thalamocortical region of the brain. The explicit
form of their equation is
\begin{equation}\label{e1}
\hat{O}_R\phi _{a}(t)=S[V_{a}(t)]~,
\end{equation}
where
\begin{equation}\label{e1bis}
\hat{O}_R=\left(\frac{1}{\nu _{a}^2}\frac{d^2}{dt^2}+\frac{2}{\nu
_{a}}\frac{d}{dt}+1-r_{a}^2 \nabla ^2\right)~,
\end{equation}
where $\nu _{a}=v_{a}/r_{a}$, $r_a$ is the mean range of axons $a$,
and $V_{a}=\sum _b V_{ab}$ is a so-called cell body potential which
results from the filtered dendritic tree inputs. Robinson has used
the experimental parameters in this equation for the processing of
the experimental data. In the following we concentrate on a
particular mathematical aspect of this equation and refer the reader
to the works of Robinson's group for more details concerning this
equation.

\medskip

\subsection{The homogeneous equation}

\medskip

\noindent We treat first the homogeneous case, i.e., $S=0$ and we
discard the subindexes as being related to the phenomenology not to
the mathematics. Let us employ the change of variable $z=ax+by-ct$
(see, e.g., \cite{est}), which is a traveling coordinate in 2+1
dimensions. This is justified because it was noticed by Wilson and
Cowan \cite{wc73} that distinct anatomical regions of cerebral
cortex and of thalamic nuclei are functionally two-dimensional
although extending to three spatial coordinates is trivial. We have
the following rescalings of functions: $\phi _t=-c\phi _z$, $\phi
_{tt}=v^2\phi _{zz}$, $\phi _{xx}=a^2 \phi _{zz}$, $\phi _{yy}=b^2
\phi _{zz}$. Then, we get the ordinary differential equation
corresponding to the damped wave equation in the following form
\begin{equation}\label{e2}
\hat{O}_{R,z}\phi\equiv\left(\frac{d^2}{dz^2}-2\mu\frac{d}{dz}+\mu^2\right)\phi
=\alpha ^2 \phi ~,
\end{equation}
where
\begin{equation}\label{e3}
\mu= \frac{\nu c}{c^2-\nu ^2r^2(a^2+b^2)}~, \qquad \alpha ^2
=\frac{\nu ^4r^2(a^2+b^2)}{[c^2-\nu ^2r^2(a^2+b^2)]^2}~.
\end{equation}
The simple damped oscillator equation (\ref{e2}) can be easily
factorized
\begin{equation}\label{e4}
L_{\mu}^2\phi\equiv\left(\frac{d}{dz}-\mu\right)\left(\frac{d}{dz}-\mu\right)\phi
=\alpha ^2 \phi ~.
\end{equation}
The case $c^2<\nu ^2r^2(a^2+b^2)$ implies $\mu <0$ and the general
solution of (\ref{e2}) can be written
\begin{equation}\label{e5}
\phi(z)=e^{\mu z} (Ae^{\alpha z} +Be^{-\alpha z})~.
\end{equation}
The opposite case $c^2>\nu ^2r^2(a^2+b^2)$ will lead to only a
change of sign in front of $\mu$ in all formulas henceforth, whereas
the case $c^2=\nu ^2r^2(a^2+b^2)$ will be considered as nonphysical.
The non-uniqueness of the factorization of second-order differential
operators has been exploited in a previous paper \cite{rr98} on the
example of the Newton classical damped oscillator, i.e.,
\begin{equation}\label{e6}
\hat{N}y\equiv \left(\frac{d^2}{dt^2}+2\beta\frac{d}{dt}+\omega
_{0}^{2}\right)y=0~,
\end{equation}
which is similar to the equation (\ref{e2}), unless the coefficient
$2\beta$ is the friction constant per unit mass,
$\omega _{0}$ is the natural frequency of the oscillator, and the
independent variable is just time not the traveling variable.
Proceeding along the lines of \cite{rr98}, one can search for the
most general isospectral factorization
\begin{equation}\label{e7}
(D_z+f(z))(D_z+g(z))\phi=\alpha ^2 \phi~.
\end{equation}
After simple algebraic manipulations one finds the conditions
$f+g=-2\mu$ and $dg/dz+fg=\mu ^2$ having as general solution
$f_{\lambda}=\frac{\lambda}{\lambda z+1}-\mu$, whereas $f_0=-\mu$ is
only a particular solution. Using the general solution $f_{\lambda}$
we get
\begin{equation}\label{e8}
\hat{A}_{+\lambda}\hat{A}_{-\lambda}\phi
\equiv\left(D_z+\frac{\lambda}{\lambda
z+1}-\mu\right)\left(D_z-\frac{\lambda}{\lambda
z+1}-\mu\right)\phi=\alpha ^2 \phi~.
\end{equation}
This equation does not provide anything new since it is just
equation (\ref{e3}). However, a different operator, which is a
supersymmetric partner of (\ref{e8}) is obtained by applying the
factorizing $\lambda$-dependent operators in reversed order
\begin{equation}\label{e9}
\hat{A}_{-\lambda}\hat{A}_{+\lambda}\tilde{\phi}
\equiv\left(D_z-\frac{\lambda}{\lambda
z+1}-\mu\right)\left(D_z+\frac{\lambda}{\lambda
z+1}-\mu\right)\tilde{\phi}=\alpha ^2 \tilde{\phi}~.
\end{equation}
The latter equation can be written as follows
\begin{equation}\label{e10}
\hat{\tilde{O}}_{\lambda}\tilde{\phi}\equiv
\left(\frac{d^2}{dz^2}-2\mu\frac{d}{dz}+\mu ^{2}-\alpha^2-
\frac{\lambda ^2}{(\lambda z+1)^2}\right)\tilde{\phi}=0~,
\end{equation}
or
\begin{equation}\label{e11}
 \left(\frac{d^2}{dz^2}-2\mu\frac{d}{dz}+\omega ^2(z)\right)\tilde{\phi}=0~,
\end{equation}
where
\begin{equation}\label{e12}
\omega ^2(z)=\mu ^{2}-\alpha^2- \frac{\lambda ^2}{(\lambda z+1)^2}
\end{equation}
is a sort of parametric angular frequency with respect to the
traveling coordinate.

 This new second-order linear damping equation contains the
additional last term with respect to its initial partner, which may
be thought of as the Darboux transform part of the frequency
\cite{D}. $Z_{\lambda}=1/\lambda$ occurs as a new traveling scale in
the damped wave problem and acts as a modulation scale. If this
traveling scale is infinite, the
ordinary damped wave problem is recovered. 
The $\tilde{\phi}$ modes can be obtained from the $\phi$ modes by
operatorial means \cite{rr98}.

Eliminating the first derivative term in the parametric damped
oscillator equation (\ref{e11}) one can get the following Bessel
equation
\begin{equation}\label{e13}
 \frac{d^2u}{dx^2}-\left(\frac{n^2-\frac{1}{4}}{x^2}+\beta ^2\right)u=0~,
\end{equation}
where $x=z+1/\lambda$, $n^2=5/4$, and $\beta =i\alpha$. Using the
latter equation, the general solution of equation~(\ref{e11}) can be
written in terms of the modified Bessel functions
\begin{equation}\label{e14}
\tilde{\phi}=(z+1/\lambda)^{1/2}[C_1I_{\sqrt{5}/2}(\alpha
(z+1/\lambda))+C_2I_{-\sqrt{5}/2}(\alpha (z+1/\lambda))]e^{\mu z}~.
\end{equation}
What could be a right interpretation of the supersymmetric partner
equation (\ref{e9}) ? Since the solutions are modified Bessel
functions, we consider this equation as a diffusion equation with a
diffusion coefficient depending on the traveling coordinate.
Noticing that the velocity in the traveling variable of this
diffusion is the same as the velocity of the neuronal pulses we
identify it with the diffusion of various molecules, mostly
hormones, in the extracellular space (ECS) of the brain, which is
known to be necessary for chemical signaling and for neurons and
glia to access nutrients and therapeutics occupying as much as 20 \%
of total brain volume {\em in vivo} \cite{ECS}.

\medskip

\subsection{The nonhomogeneous equation}

\medskip

The source term $S$ in Robinson's equation (\ref{e1}) is a sigmoidal
firing function, which despite corresponding to a realistic case led
him to work out extensive numerical analyses. Analytic results have
been obtained recently by Troy and Shusterman \cite{TS} by using a
source term comprising a combination of discontinuous exponential
coupling rate functions and Heaviside firing rate functions. In
addition, Brackley and Turner \cite{BT} incorporated fluctuating
firing thresholds about a mean value as a source of noisy behavior
\cite{BT}.

The procedure of Troy and Shusterman can be applied for the
parametric damped oscillator equation as well as for the Bessel
diffusion equation obtained herein in the realm of Robinson's brain
wave equation with the difference that the method of variation of
parameters should be employed. The detailed mathematical analysis is
left for a future work.

\section*{Acknowledgment}
This work was partially supported by the CONACyT Project 46980.


\end{document}